\documentclass[aps,prl,reprint,superscriptaddress]{revtex4-1}
\usepackage{graphicx}
\usepackage{amsmath}

\begin{document}


\title{Cavity sideband cooling of the Josephson phase}

\author{J. Hammer}
\affiliation{Institute for Experimental and Applied Physics, University of Regensburg, D-93040 Regensburg, Germany.}
\author{M. Aprili}
\email[]{aprili@lps.u-psud.fr}
\affiliation{Laboratoire de Physique des Solides, UMR8502-CNRS, University Paris-Sud, 91405 Orsay Cedex, France.}
\author{I. Petkovi\'{c}}
\affiliation{Laboratoire de Physique des Solides, UMR8502-CNRS, University Paris-Sud, 91405 Orsay Cedex, France.}

\date{\today}

\begin{abstract}
An extended Josephson junction intrinsically couples the superconducting current to the microwave cavity in the insulating barrier. We demonstrate that this coupling produces sidebands in the microwave cavity resonances of the junction. By measuring the switching current distribution, we show that microwave radiation at sidebands brings the Josephson phase out of equilibrium. In particular, the effective phase temperature is  reduced or enhanced through anti-Stokes and Stokes scattering, respectively. Phase cooling and heating both increase with microwave power.

 \end{abstract}
%
%

\maketitle

The radiation pressure is used for cooling atoms, ions and optomechanical devices. In a Fabry-P$\acute{\text{e}}$rot (FP) interferometer in which one of the two mirrors vibrates, the Brownian motion of the vibrating mirror and hence its effective temperature can surprisingly be lowered by increasing the power of light \cite{braginskii1,braginskii2,arcizet}. There is a straightforward analogy with a Josephson junction irradiated with microwave photons, where the phase difference between the wavefunctions of two superconductors, the Josephson phase, takes the role of the mirror position. In this Letter we show that the microwave field acts on the Josephson phase as the radiation pressure does on a vibrating mirror. Specifically, when coupled with a high quality microwave cavity, the Josephson junction generates sideband resonances for each cavity mode. Phase heating or cooling is achieved by microwave radiation at these sidebands.

\begin{figure}[t]
		\includegraphics[width=8.6cm]{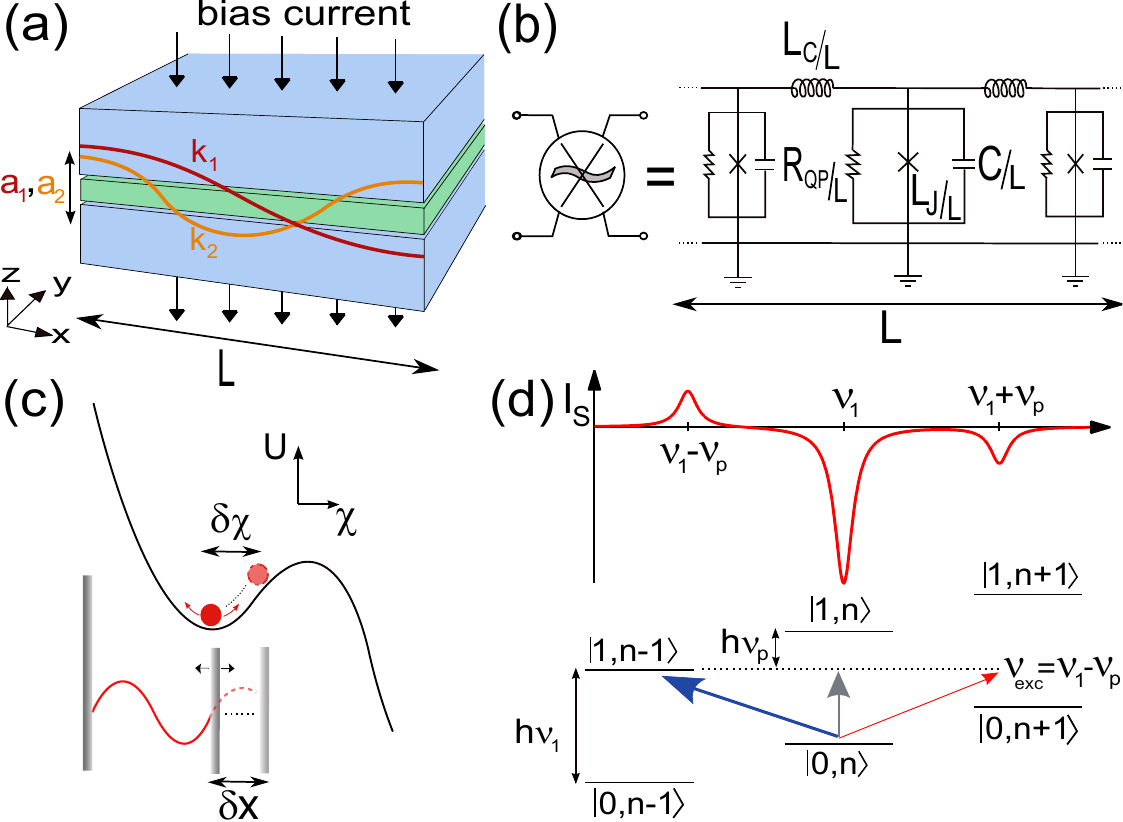}
			\caption{\label{fig:FIG1}(a) Sketch of an extended Josephson junction supporting resonant electromagnetic modes $k_n  = n \cdot {\raise0.5ex\hbox{$\scriptstyle \pi $}
\kern-0.1em/\kern-0.15em
\lower0.25ex\hbox{$\scriptstyle L$}}$ with intra-cavity field amplitudes $a_n$ across the insulating barrier. (b) Electromagnetic transmission line model of an extended Josephson junction. (c) Analogy between cooling of phase oscillations in the tilted washboard potential and opto-mechanical cooling in a Fabry-Perot cavity. In junctions with high finesse, microwaves act similarly on phase dynamics as the radiation pressure on a moving mirror. (d) Schematic of Stokes and anti-Stokes processes due to sideband radiation.
}\end{figure}

A mm-scale Josephson junction, formed by two superconducting electrodes separated by an insulating barrier, is in fact a microwave cavity and a Josephson resonator. An artistic view of a Josephson junction of length L  and its electromagnetic modes is presented in Fig.~\ref{fig:FIG1}(a). Electrically, a distributed Josephson  junction is analogous to a transmission line as depicted in Fig.~\ref{fig:FIG1}(b). The geometrical inductance, $L_C$, and the capacitance, $C$, define the microwave cavity. While the Josephson resonator is formed by $L_J$ and C, the single electron and Cooper pair tunneling give rise to the dissipative, $R_{QP}$, and the inductive, $L_J  = {\hbar  \mathord{\left/
 {\vphantom {\hbar  {2eI_c \cos \chi}}} \right.
 \kern-\nulldelimiterspace} {2eI_c \cos \chi }}$ transport through the junction, respectively \cite{likharev}. Here $I_c$ is the junction critical current and $\chi$ the phase difference of the wavefunctions between the two superconductors with no excited modes in the cavity. The cavity resonances are parametrically coupled to  the oscillations of the Josephson phase, $\chi$. If the intracavity field does not follow adiabatically the phase oscillations because of the high finesse of the cavity,  the delay between the motion of the phase and the photo-assisted phase dynamics causes an additional friction\cite{arcizet} (the analogy with a FP interferometer is illustrated in Fig.1(c)). First we address the coupling between the Josephson phase and the cavity modes and then we focus on photo-assisted phase dynamics.%
 
In the limit of  small microwave power $P_{RF}$ and zero external DC-voltage and applied magnetic field, the Hamiltonian of the junction obtained from the Josephson and the Maxwell relations can be written as $H_T=H_J+H_C+H_{int}$ given by \cite{fistul}
	\begin{subequations}
		\begin{align}
		H_J  &= \left[ \frac{1}{{8\pi^2\nu _p^2 }}\left( {\frac{{d\chi }}{{dt}}} \right)^2  + U(\chi )\right]   \\
	   H_C  &= \left[\frac{1}{{16\pi^2\nu _p^2 }}\left( {\frac{{dA_1 }}{{dt}}} \right)^2  + \frac{{\nu _1^2 }}{{4\nu _p^2 }}A_1^2\right]    \\
	   	   H_{{\mathop{\rm int}} }  &=  - E_J\frac{\chi A_1^2 }{4}.
	\end{align}
	\end{subequations}
Here $H_J$ is the Hamiltonian of the Josephson junction, $H_C$ is that of the cavity and $H_{int}$ describes the interaction between the cavity and the Josephson current. The effective potential, $U(\chi)$ is given by $U(\chi) = 1 - \cos (\chi ) - (I_{in}/I_c) \,\chi $, where $I_{in}$ is the bias current,  $\nu _p$ is  the plasma frequency and  $E_J  = {\textstyle{\hbar  \over {2e}}}I_c $ is the Josephson energy. For simplicity we consider the first mode $A_1$ only with $A_1 (t) = a_1 e^{i 2\pi \nu _1 t} $ where $a_n  = {{2eV_n } \mathord{\left/
 {\vphantom {{2eV_n } {h \nu _n }}} \right.
 \kern-\nulldelimiterspace} {h \nu _n }}$ is the intracavity field amplitude of the corresponding mode.%

If there are no photons present in the cavity, i.e. classically $A_1=0$ and hence $H_C=0$ and $H_{int}=0$, the dynamics of the phase difference, $\chi$, is equivalent to that of a particle of mass $M=(2e/h)^2C$  in a washboard like potential $U(\chi)$ \cite{tinkham} as shown in Fig.~\ref{fig:FIG1}(c). The tilt of the washboard potential is given by the current bias, $I_{in}$. For $I_{in}<I_c$ thermal fluctuations help the Josephson phase escape from the washboard minimum. This results in a switching current to the dissipative state, $I_s$, smaller than $I_c$ and more importantly in a thermal distribution of the switching current \cite{fulton} so that switching is a direct measure of the phase temperature \cite{MQT}. %

The interaction term $H_{int}$ arises from the non-linear current-phase Josephson relation and it is formally equivalent to the radiation pressure in a FP interferometer \cite{fabre} except that here the mechanical coordinate is the phase. This is shown in Fig.~\ref{fig:FIG1}(c) where the Josephson oscillations replace the mirror oscillations of the FP  interferometer. When coupled dynamics are ignored the effect of $A_1$ on $I_c$ is trivial such that $I_c(A_1)= I_c(1- A_1^2)$ and simply shows the standard reduction of $I_c$ by microwaves\cite{shapiro}. However for $\nu_p$ larger than the cavity bandwidth $\nu_B$, a static approach is insufficient to describe the Josephson dynamics which is instead strongly correlated to that of the electromagnetic field in the cavity. To investigate the coupled dynamics we have to consider at the same time the cavity storage time, the Josephson induced dephasing due to the electromagnetic field in the cavity and damping of the Josephson oscillations.

The classical dynamical equation for the intracavity field amplitude is \cite{fabre}
\begin{equation}
			\begin{aligned}
	   \frac{{da_1 }}{{dt}} &=  -2\pi \left( {\nu _B  - i\nu _D } \right)a_1  + f a_{in}     \\
	  {\rm with}\quad \nu _D  &=  - \Delta \nu  + \frac{{\nu _p^2 }}{{2\nu _1 }}\left( {\chi \left( t \right) - \left\langle \chi  \right\rangle } \right),
	\end{aligned}
\end{equation}
\noindent where $\nu_D$ is the detuning frequency, $f$ a geometrical parameter and $a_{in}$ the power injected into the cavity. In the expression for $\nu_D$ the first term accounts for the detuning of the microwave radiation from the cavity mode (i.e. $\Delta\nu = \nu- \nu_1$) while the second term arises from the coupling to the phase and has both a dynamical $\chi(t)$ and  static component $\left\langle \chi  \right\rangle$. Physically, the Josephson oscillations change the propagation  in the cavity through the Josephson inductance that modulates the microwave impedance of the transmission line. This produces frequency and amplitude modulation of the cavity mode. Amplitude modulation gives sidebands \cite{schliesser} in the cavity resonance as described in Fig.~\ref{fig:FIG1}(d).

The back action of the intracavity field on the Josephson phase results in an effective force acting on the phase that actually produces damping and also changes the eigenfrequency of the Josephson oscillator. Although it is natural to consider that the microwave impedance of the environment affects the phase damping \cite{graber,turlot}, coupled phase and cavity dynamics lead to an active damping which is proportional to the microwave power. By analogy with calculations for a FP interferometer with a mobile mirror\cite{arcizet2}, we find $\Gamma _{eff} = \Gamma _J \left({1 - 2Q_J {\mathop{\rm Im}\nolimits} \left\{ {\xi (\nu )} \right\}} \right)$ and $ \nu _{p,eff} = \nu _p \left( {1 + {\mathop{\rm Re}\nolimits} \left\{ {\xi (\nu )} \right\}} \right)$ with $\xi (\nu ) \propto {{P_{RF} } \mathord{\left/
 {\vphantom {{P_{RF} } {\left[ {\left( {1 - i\;{{\nu _p } \mathord{\left/
 {\vphantom {{\nu _p } {\nu _B }}} \right.
 \kern-\nulldelimiterspace} {\nu _B }}} \right)^2  + \left( {{\nu  \mathord{\left/
 {\vphantom {\nu  {\nu _B }}} \right.
 \kern-\nulldelimiterspace} {\nu _B }}} \right)^2 } \right]}}} \right.
 \kern-\nulldelimiterspace} {\left[ {\left( {1 - i\;{{\nu _p } \mathord{\left/
 {\vphantom {{\nu _p } {\nu _B }}} \right.
 \kern-\nulldelimiterspace} {\nu _B }}} \right)^2  + \left( {{\nu  \mathord{\left/
 {\vphantom {\nu  {\nu _B }}} \right.
 \kern-\nulldelimiterspace} {\nu _B }}} \right)^2 } \right]}}$. Where $\Gamma_J {\rm  = }{{\rm 1} \mathord{\left/ {\vphantom {{\rm 1} {{\rm R}_{QP} {\rm C}}}} \right. \kern-\nulldelimiterspace} {{\rm R}_{QP} {\rm C}}} $ is the damping parameter and $Q_J= 2 \pi \nu_p/ \Gamma_J$ the Josephson quality factor. As a consequence, the amplitude of the Josephson oscillations is either damped (negative-detuning) or enhanced (positive-detuning) at the sidebands corresponding to effective cooling or heating, respectively. The equipartition theorem links the "phase temperature" to its oscillation amplitude.%

\begin{figure}[t]
		\includegraphics[width=8.6cm]{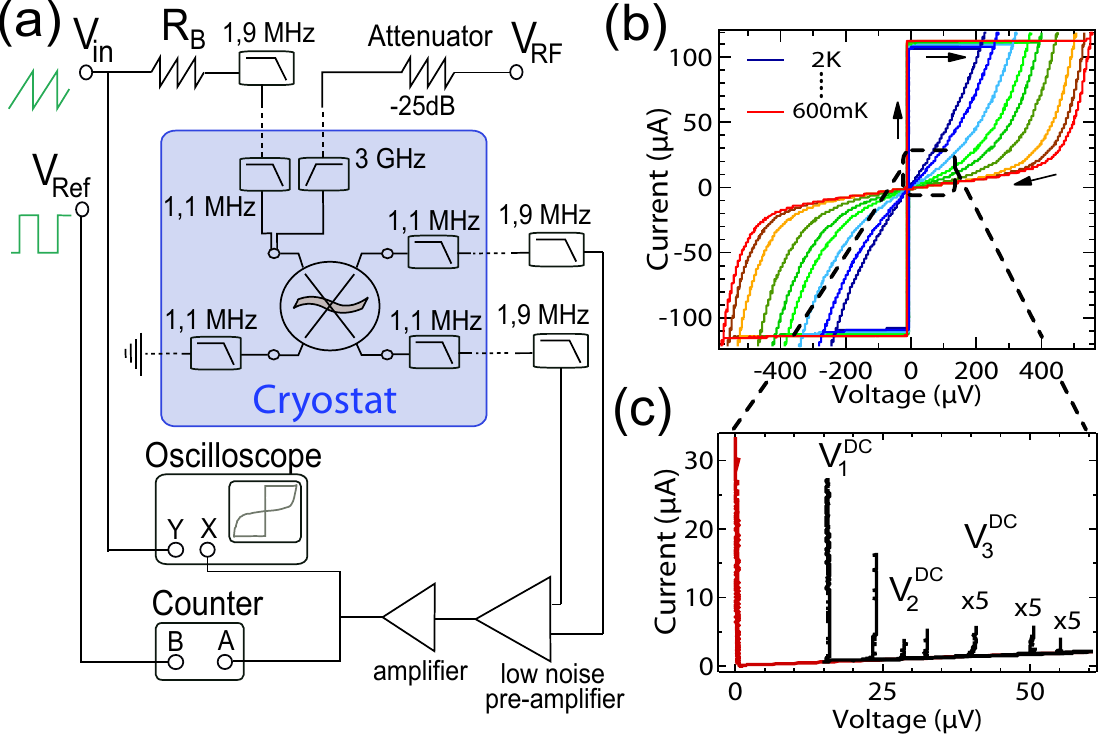}
			\caption{\label{fig:FIG2}(a) Schematic of switching current measurement setup:  the junction is current biased with  a 77Hz sawtooth ramp. The switching current probability distribution $P(I_S)$ is measured using a counter. A separate line allows microwave excitation. (a) I-V curves in a temperature range from 2K (deep blue) to 600 mK (red). (c) Detail from (b): Fiske resonances at 600 mK, corresponding to frequencies $\nu _1  = 7.0\;{\rm GHz,}\;\nu _2  = 15.0\;{\rm GHz,}\;...$ according to $V_n^{DC}  = {\raise0.5ex\hbox{$\scriptstyle h$}\kern-0.1em/\kern-0.15em
\lower0.25ex\hbox{$\scriptstyle {2e}$}} \cdot \nu _n $. Each resonance is measured for the applied in plane magnetic field that maximizes its amplitude.
}\end{figure}

We use  Nb/Al/Al$_2$O$_3$/PdNi/Nb and  Nb/Al/Al$_2$O$_3$/Nb Josephson junctions in a cross-strip geometry \cite{petkovic}. The  junction area is $740 \mu \rm{m}$ by $740 \mu \rm{m}$. Both junction types show sideband cooling and heating. The thin PdNi layer is added to reduce the critical current and the Josephson quality factor in order to make switching measurements easier at low temperature. All the data reported here have been taken in Nb/Al/Al$_2$O$_3$/PdNi/Nb junctions. The circuit used to measure the current-voltage (I-V)  characteristics  and the switching current  distribution is reported in Fig.~\ref{fig:FIG2}(a). The I-V characteristics  at temperatures between $2$ K and $600$ mK are shown in Fig.~\ref{fig:FIG2}(b). The junctions are strongly underdamped, i.e. the damping parameter   $\Gamma _J <<  \nu_p$. The lowest temperature is measured using  an Al/Al$_2$O$_3$/Al tunnel junction as a thermometer. When a finite DC-voltage appears across the junction, the cavity modes are resonantly excited at $V_n^{DC}  = \frac{h}{2e} \nu _n $, and mix with the AC-Josephson current giving rise to finite DC-resonances (Fiske steps) \cite{coon}. This is shown in Fig.~\ref{fig:FIG2}(c). The modes labeled $V_n^{DC}$ correspond to modes $k_n$, with $n=1,2$ and $3$. The frequencies of the resonant cavity modes are $\nu _n  = {\raise0.7ex\hbox{${c_s }$} \!\mathord{\left/
 {\vphantom {{c_s } {2\pi }}}\right.\kern-\nulldelimiterspace}
\!\lower0.7ex\hbox{${2\pi }$}} \cdot k_n $ with $k_n  = {{n\,\pi } \mathord{\left/
 {\vphantom {{n\,\pi } L}} \right.
 \kern-\nulldelimiterspace} L}$ and $c_s$ is the Swihart velocity \cite{swihart}. The first mode occurs at $\nu_1\sim7$ GHz. Higher symmetry modes are also visible. From the dispersion of these modes \cite{likharev} we get the Swihart velocity ${\rm c}_{\rm s} {\rm = 1.1} \cdot {\rm 10}^{\rm 7}$m/s, the junction capacitance C = 30nF and then the plasma frequency at zero current bias   $\nu _p  ={\raise0.5ex\hbox{$\scriptstyle 1$}
\kern-0.1em/\kern-0.15em
\lower0.25ex\hbox{$\scriptstyle {2\pi }$}} \sqrt {{{2eI_c } \mathord{\left/
 {\vphantom {{2eI_c } {\hbar C}}} \right.
 \kern-\nulldelimiterspace} {\hbar C}}} =550 $MHz $\ll \nu_1$.

\begin{figure}[b]
		\includegraphics[width=8.6cm]{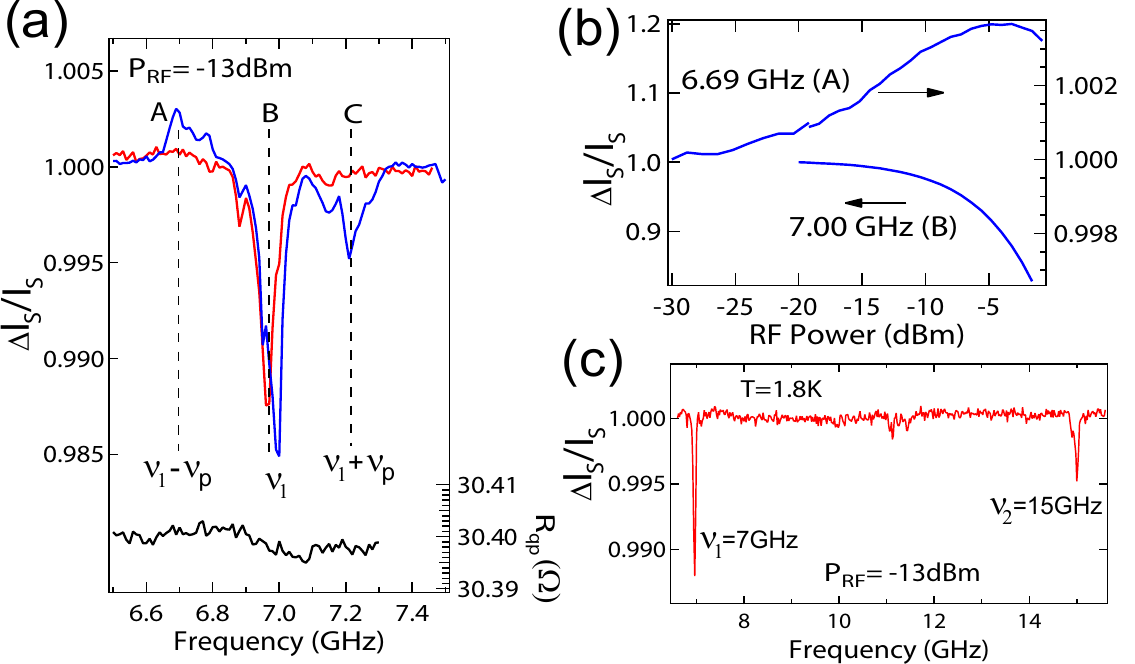}
			\caption{\label{fig:FIG3}a) Switching current as function of microwave frequency at T=2K (blue) and T=600mK (red). At low temperatures sidebands occur, detuned about the plasma frequency $\nu_p$. The quasiparticle resistance $R_{qp}$ (black curve below) at 600mK does not change with the RF-frequency indicating no change in the junction electron temperature. b) Switching current as a function of  RF-power from $-30$ dBm to $-2$ dBm at 600mK  taken at the resonance frequency (B): $\nu=\nu_1$ and at the anti-Stokes frequency (A): $\nu=\nu_1-\nu_p$ . c) Whole spectrum from $6.5$ GHz to $16$ GHz  measured at $-13$ dBm and T=2K. The first two resonances at $\nu _1  = 7\;{\rm GHz,}\;\nu _2  = 15\;{\rm GHz}$, coincide precisely with the Fiske steps in Fig.~\ref{fig:FIG2}(c).}
\end{figure}

In the following we focus on the photo-assisted dissipationless current. This is obtained by microwave radiation through the RF-excitation line (Fig.~\ref{fig:FIG2}(c)). The high-pass filter ensures negligible microwave power at the Josephson plasma frequency. Josephson spectroscopy of the cavity is shown in Fig.~\ref{fig:FIG3}(c) where the Josephson critical current is measured as a function of the microwave frequency. When the cavity is externally pumped, the Josephson current decreases. These modes correspond precisely to those measured resonantly by the AC-Josephson effect and reported in Fig.~\ref{fig:FIG2}(c). We obtain the cavity quality factor of about 200 independent of temperature below 2K. The frequency of the cavity modes is much larger than $\nu_p$ and hence the dielectric losses in the insulating barrier limit the value of  the cavity quality factor rather than the single particle tunneling.

The coupled dynamics appear in the switching current at low temperature (below $1.5$K) when the Josephson quality factor is large. Moreover, at the lowest temperature ($600$ mK), the phase relaxation time $\tau_{\chi} \sim 200\mu$s \cite{petkovic} is much longer than the electron-phonon relaxation time $\tau_{e-ph} \sim 3\mu$s \cite{latempa}, ensuring good insulation from the thermal bath. In Fig.~\ref{fig:FIG3}(a) sidebands at $\nu_1\pm\nu_p$ are revealed through the frequency dependence of the switching current. The imposed current bias changes the washboard potential and hence the plasma frequency is reduced by about a factor 2 at switching. Higher damping enhances the switching current while lower damping decreases it. Note that the quasiparticle resistance is frequency independent as shown in Fig.~\ref{fig:FIG3}(a). The quasiparticle resistance $R_{QP}$ is thermally activated and it is therefore an ideal thermometer of the electron bath. This stresses out that microwave induced phase cooling and heating are out-of-equilibrium phenomena. Fig.~\ref{fig:FIG3}(b) shows the change of the switching current as a function of the microwave power at the cavity resonance, $\nu_1$, and the anti-Stokes, $\nu_1-\nu_p$, frequency. As expected, higher microwave power at $\nu=\nu_1-\nu_p$ produces higher phase damping  and hence larger switching currents while the higher microwave power in the main cavity mode suppresses the critical current further. Throughout the paper $P_{RF}$ refers to the source output. A quantitative estimation of the $f$ parameter in Eq. (3) is beyond the scope of this work.

Although sideband cooling is substantially a classical mechanism \cite{kippenberg}, it is pedagogical to quickly illustrate its quantum analogue \cite{wilson-rae,marquardt}. When photons of energy $\nu_1-\nu_p$ reach the junction, they combine with Josephson plasmons through anti-Stokes scattering: the occupation of the cavity increases, while the occupation of the Josephson oscillator decreases. As the escape of a photon from the cavity does not change the amplitude of the Josephson oscillator this scattering leads to cooling. Effective heating corresponds to photons of energy $\nu_1+\nu_p$ that add photons to the cavity and plasmons to the junctions (i.e. Stokes scattering). This mechanism is illustrated in Fig.~\ref{fig:FIG1}(d).

\begin{figure}[]
		\includegraphics[width=8.6cm]{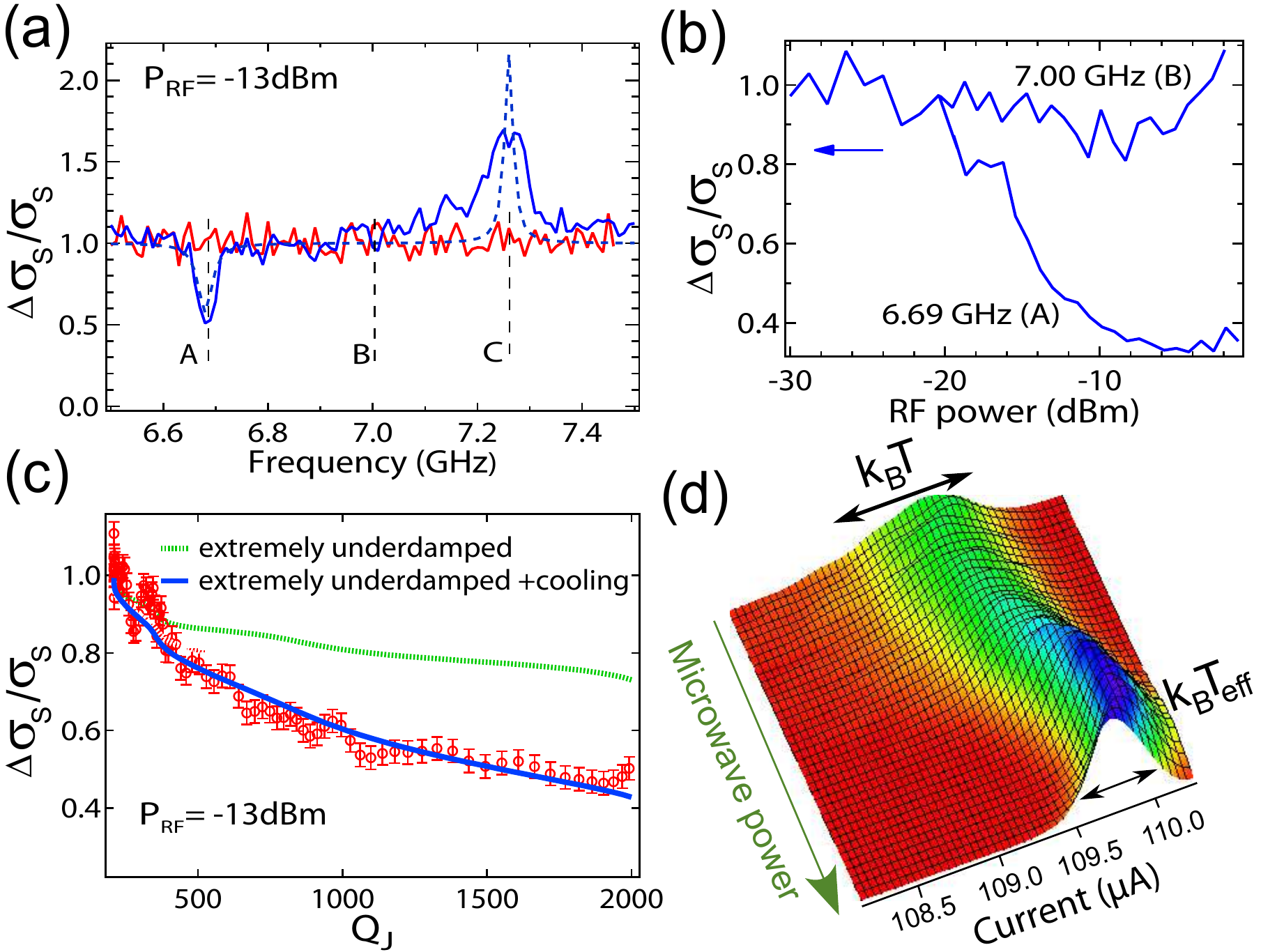}
			\caption{\label{fig:FIG4}a) Histogram width $\sigma_S$ measured at T=2 K (red) and T=600 mK (blue) as function of microwave frequency with theoretical fit (dashed blue) obtained from the expression of the RF-induced effective damping described in the text. Cooling of the Josephson phase occurs at the Stokes line (A), whereas heating is observed at the Stokes frequency (C). At resonance $\nu   = \nu _1$ there is no effect on $\rm{T_{eff}}$. b) Relative change of $\sigma_S$ at $\nu=6.69$ GHz and $\nu=7$ GHz as a function of the microwave power from $-30$ dBm to $-2$ dBm at 600 mK. c) Relative change in $\sigma_S$ due to sideband cooling as a function of the junction quality factor.  Data are obtained by radiation at the anti-Stokes frequency $\nu=6.69$ GHz by decreasing the junction temperature from 2 K to 600 mK and normalization by $\sigma_S$ at 2K without  RF radiation. Dashed lines are theoretical predictions with (blue) and without (green) microwave radiation. d) Switching current histograms for increasing RF-power at $\nu=6.69$ GHz. Cooling corresponds to narrower histograms at higher power.}
			\end{figure}

At $T \ll T_c$, the switching current is only weakly dependent on  temperature; however the "phase temperature" is experimentally discernable from the  width $\sigma_s$ of the switching histograms. Fig.~\ref{fig:FIG4}(a) shows the variations of $\sigma_s$ as a function of microwave frequency around the first cavity mode (similar behavior is observed for higher modes). For the anti-Stokes and Stokes frequencies (labelled A and C), $\sigma_s$ is strongly reduced or enhanced. This confirms cooling for negative sideband detuning and heating for positive sideband detuning. No changes in the phase temperature are observed at zero detuning independent of the microwave power (Fig.~\ref{fig:FIG4}(b)). Cooling power instead increases with microwave power. The evolution of the switching histogram for irradiation at he anti-Stokes frequency is plotted in Fig.~\ref{fig:FIG4}(d). We have achieved a minimum phase temperature of 320 mK (at 600 mK cryostat temperature) which is consistent with the measured maximum increase of the switching current under microwave irradiation at  $\nu_1-\nu_p$. The shape of the histograms verifies the current bias dependence expected from the Kramers escape rate \cite{fulton}.%

The phase temperature is given by ${{T_{eff} } \mathord{\left/
 {\vphantom {{T_{eff} } T}} \right.
 \kern-\nulldelimiterspace} T} = {\Gamma  \mathord{\left/
 {\vphantom {\Gamma  {\Gamma _{eff} }}} \right.
 \kern-\nulldelimiterspace} {\Gamma _{eff} }}$, while in the context of Kramers theory $\sigma_s$ is proportional to $\Gamma_J ^{{\raise0.5ex\hbox{$\scriptstyle 2$}
\kern-0.1em/\kern-0.15em
\lower0.25ex\hbox{$\scriptstyle 3$}}} $
and hence to $T^{{\raise0.5ex\hbox{$\scriptstyle 2$}
\kern-0.1em/\kern-0.15em
\lower0.25ex\hbox{$\scriptstyle 3$}}} $. The blue dotted curve in Fig.~\ref{fig:FIG4}(a) is the theoretical expectation leaving the microwave power as fitting parameter. The effective damping is temperature dependent through the Josephson quality factor. This is presented in Fig.~\ref{fig:FIG4}(c) where the width of the switching histograms is shown for anti-Stokes cooling. Lowering the temperature increases the quality factor and hence the effective damping (see equation for $\Gamma_{eff}$ given above). The dotted green curve in Fig.~\ref{fig:FIG4}(c) is the theoretical expectation for $\sigma_s$ taking into account only thermal fluctuations, the solid blue curve is obtained adding sideband cooling.

In conclusion, we have observed sideband cooling and heating of the Josephson phase by microwave excitation. There is a direct mapping with cooling and heating of a mechanical oscillator in a Fabry-Perot interferometer by radiation pressure. Due to the strong coupling between the phase and the photon dynamics, extended Josephson junctions offer new perspectives  in the merging field of combined quantum transport and quantum optics. Moreover, it is instructive to point out that some technical aspects of a large-scale interferometer may be investigated using a simple Josephson junction.

\begin{acknowledgments}
We thank I. Favero, B. Reulet and C. Strunk for valuable discussions. We are indebted to  J. Gabelli who brought  to our attention the experiments on optomechanical cooling. J.H. acknowledges support through the Max Weber-Program and the Erasmus Program.
\end{acknowledgments}


\begin{thebibliography}{1}
\bibitem{braginskii1} V.B. Braginskii and A.B. Manukin, JETP \textbf{25}, 653 (1967)
\bibitem{braginskii2} V.B. Braginskii and S.P. Vyatchanin, Phys. Lett. A \textbf{293}, 228 (2002)
\bibitem{arcizet} O. Arcizet \textit{et al.}, Nature( London) \textbf{444}, 71 (2006)
\bibitem{likharev} K.K. Likharev, Rev. Mod. Phys. \textbf{51}, 101 (1979)
\bibitem{fistul} M.V. Fistul and A.V. Ustinov, Phys. Rev. B \textbf{138}, 214506 (2007)
\bibitem{tinkham} M. Tinkham, \textit{Introduction to Superconductivity} (Mc-Graw-Hill, New York, 1996), 2nd ed.
\bibitem{fulton} T.A. Fulton and L.N. Dunkelberger, Phys. Rev. B \textbf{9}, 4760 (1974)
\bibitem{MQT} In our experiment quantum tunneling of the phase  can be disregarded as it appears below $T^*  = {{h \nu _p } \mathord{\left/
 {\vphantom {{h \nu _p } {2\pi k_B \;( \approx 15{\rm mK})}}} \right.
 \kern-\nulldelimiterspace} {2\pi k_B \; \approx 25{\rm mK}}}$.
\bibitem{fabre} C. Fabre \textit{et al.}, Phys. Rev. A \textbf{49}, 1337 (1994)
\bibitem{shapiro} S. Shapiro, Phys. Rev. Lett. \textbf{11}, 80 (1963)
\bibitem{schliesser} A. Schliesser \textit{et al.}, Nature Physics \textbf{4}, 415 (2008)
\bibitem{graber} H. Graber and S. Linkwitz, Phys. Rev. A. \textbf{37}, 963 (1988)
\bibitem{turlot} E. Turlot \textit{et al.}, Phys. Rev. Lett. \textbf{62}, 1788 (1989)
\bibitem{arcizet2} O. Arcizet \textit{et al.}, Phys. Rev. A. \textbf{73}, 033819 (2006)
\bibitem{petkovic} I. Petkovic and M. Aprili, Phys. Rev. Lett. \textbf{102}, 157003 (2009)
\bibitem{coon} D.D. Coon and M.D. Fiske, Phys. Rev. \textbf{432}, A744 (1965)
\bibitem{swihart} J.C. Swihart, J. of Appl. Phys. \textbf{32}, 461 (1961)
\bibitem{latempa} R. Latempa, M. Aprili and I. Petkovic, J. Appl. Phys. \textbf{106}, 103925 (2009)
\bibitem{kippenberg} T.J. Kippenberg \textit{et al.}, Phys. Rev. Lett. \textbf{95}, 033901 (2005)
\bibitem{wilson-rae} I. Wilson-Rae, P. Zoller and A. Imamoglu, Phys. Rev. Lett. \textbf{92}, 075507 (2004)
\bibitem{marquardt} F. Marquardt \textit{et al.}, Phys. Rev. Lett. \textbf{99}, 093902 (2007)
\end{thebibliography}

\end{document}